\documentclass[aps,pre,twocolumn,superscriptaddress]{revtex4}

\usepackage[utf8x]{inputenc} 
\usepackage{ae} 
\usepackage[T1]{fontenc}
\usepackage[english]{babel}
\usepackage{amsmath}
\usepackage{amsfonts}
\usepackage{amssymb}
\usepackage{array}
\usepackage{graphics}
\usepackage{graphicx}
\usepackage{epsfig}
\usepackage{latexsym}
\usepackage{textcomp}
\usepackage{gensymb}
\usepackage{verbatim}
\usepackage{subfigure}
\usepackage[retainorgcmds]{IEEEtrantools}
\newcommand{\rd}{\mathrm{d}}
\newcommand{\nl}{\nonumber \\}
\newcommand{\mc}{\IEEEeqnarraymulticol}
\newcommand{\la}{\langle}
\newcommand{\ra}{\rangle}
\newcommand{\eps}{\varepsilon}

\newcommand{\tvec}[2]{\big(\! \begin{smallmatrix} #1 \\ #2 \end{smallmatrix} \!\big)}

\DeclareMathOperator{\sgn}{sgn}

\begin{document}

\title{Umbilical points of a non-Gaussian random field}
\date{}

\author{A.M. Turner}
  \affiliation{Institute for Theoretical Physics, Universiteit van Amsterdam, NL 1090 GL Amsterdam, The Netherlands}
\author{T.H. Beuman}
\author{V. Vitelli}
  \email{vitelli@lorentz.leidenuniv.nl}
  \affiliation{Instituut-Lorentz for Theoretical Physics, Leiden University, NL 2333 CA Leiden, The Netherlands}

\date{\today}

\begin{abstract}

Random fields in nature often have, to a good approximation, Gaussian characteristics. For such fields, the relative densities of umbilical points -- topological defects which can be classified into three types -- have certain fixed values. Phenomena described by nonlinear laws can however give rise to a non-Gaussian contribution, causing a deviation from these universal values. We consider a Gaussian field with a perturbation added to it, given by a nonlinear function of that field, and calculate the change in the relative density of monstars. This allows us not only to detect a perturbation, but to determine its size as well. This geometric approach offers an independent way of detecting non-Gaussianity, which even works in cases where the field itself cannot be probed directly.

\end{abstract}

\maketitle

Random surfaces can be characterized by a variety of statistical measures that are geometrical and topological in character. The most well-known are perhaps the Minkowski functionals and the statistics of critical points.
In this paper, we focus on a class of singular points of the surface, known as umbilics, that do not depend on how the surface is oriented in space. In order to understand the geometrical meaning of umbilical points, imagine drawing at every point on the surface the two principal directions, along which its curvature is maximal or minimal. At some locations the principal directions cannot be defined, because the curvature is the same along all directions -- these special points are called \emph{umbilics}. As we shall see, umbilical points are topological defects with an index of $\pm1/2$.   

This geometrical construct is very useful in a number of physical contexts. In statistical optics,  the surface may represent a curved wavefront that emerges when a plane wave is passed through an inhomogeneous refracting medium. In this mapping, the normals to the surface are light rays and the umbilical points correspond to the regions where the wave attains its maximal intensity. In two-dimensional elasticity or fluid flow, the surface can represent a potential function of two variables, whose second derivatives define a shear field that corresponds to the principal curvature directions of the surface. The points where the shear field vanishes are the umbilical points.

The umbilical points of a surface can be classified into three types: lemons, monstars and stars (see figure~1). A striking statistical feature of surfaces whose height fluctuates spatially like an isotropic Gaussian random field is that the densities of the three types of umbilics have fixed ratios, which are universal numbers \cite{cite_Berry, cite_Dennis2}. This property can therefore be used to test whether a given isotropic field is Gaussian; if for a given field $h$ the relative densities are found to differ from the universal values, one may immediately conclude that the field under consideration is not an isotropic Gaussian one. Crucially such a test requires only that the line field corresponding to the principal curvature directions is measurable -- the statistics of the scalar height field from which the curvature directions are derived can be probed without being directly observed.

To give an example of a case where the near-Gaussian field of interest is not directly observable, consider the phenomenon of weak gravitational lensing \cite{cite_Hoekstra}. As stipulated by the theory of general relativity, matter bends spacetime, which also affects light rays. The light from a distant galaxy for instance, does not come to us in a straight line, due to the presence of matter between that galaxy and us. As a result, we see a distorted image of the galaxy. In general, a circular object will look like an ellipse. While most of the matter in the universe is believed to be made up of dark matter which we cannot (yet) detect, the shear field \emph{can} be detected. The near-Gaussian field in this case is obtained by projecting the mass onto the sky, along the lines of sight. This is called the \emph{projected gravitational potential}. On large scales, this field is approximately Gaussian by virtue of the central limit theorem, since the projection involves summing over a lot of regions which are randomly distributed. On smaller scales however, interactions can give rise to non-Gaussian contributions. If we interpret the projected gravitational potential as a (near-Gaussian) surface, then the shear direction corresponds to the principal direction of this surface \cite{cite_Vitelli}. At the umbilical points the shear direction cannot be defined, hence the amplitude of the shear field must vanish -- at these locations in the sky, a circular light source still appears circular.

Another example of a physical process in which umbilical points can prove their usefulness is in the context of optical speckle fields. These fields arise for example when a coherent beam of light scatters from a rough surface. Since the many reflected waves become superimposed, this produces a random pattern of intensity with approximately Gaussian statistics. In this case, it is the points of circular polarization that can be identified as umbilical points. The relative densities of the various types of umbilical points have been found to match the theoretical predictions in experiments \cite{cite_Flossmann}. 
A speckle field is not always Gaussian. First, when the surface is not that rough, the superposition of the reflected waves will not be sufficiently random. Second, a light beam could be transmitted through a random medium to map out the statistics of its index of refraction

As a final example, consider a two-dimensional ferromagnet above the critical temperature, for which we measure the $z$-component of the magnetization as a function of $x$ and $y$. As long as the correlation length of the magnetization is smaller than the resolution of the measurement, many independent regions are averaged together and a Gaussian signal results. However, as one approaches the critical point, the correlation length diverges, resulting in non-Gaussianity.

Other contexts in which umbilical points can offer a window for non-Gaussianity include polarization singularities in the cosmic microwave background \cite{cite_Dennis3, cite_Huterer, cite_Vachaspati, cite_Naselsky}, topological defects in a nematic \cite{cite_Hudson, cite_Dzyaloshinskii} and a superfluid near criticality \cite{cite_Halperin,cite_Liu}.


Testing whether the three types of umbilical points occur in their prescribed ratios can thus reveal whether a non-Gaussian component is present in a given field. However, it does not provide any quantitative information on the size of the non-Gaussianity. In this paper, we address precisely this issue, by calculating how much the relative densities of umbilical points deviates from the universal values in the relation to the type and size of the perturbation. In our previous paper \cite{cite_paper1}, we employed the imbalance between the maxima and minima of a field to attack the same problem. Besides being applicable even when the field itself cannot be observed directly, the approach based on umbilics provides an additional probe, should the extrema test not be sensitive enough. As an illustration, consider the case $h(\vec{r}) = H(\vec{r}) + \eps H(\vec{r})^3$, where $H(\vec{r})$ is a Gaussian field. Since the perturbation is an odd function of $H$, the symmetry between positive and negative values of $H$ is preserved and the relative densities of maxima and minima will not differ. By contrast, a study of the umbilical points does reveal the non-Gaussianity of $h$, as we will show.

The outline of this paper is as follows. In section~\ref{sec_gsn_fields}, we briefly review the basic properties of Gaussian fields while in section~\ref{sec_umb}, we introduce the necessary geometric concepts concerning umbilical points. In sections~\ref{sec_gen_func} through \ref{sec_prob_distr} the various steps and concepts that are needed for this calculation are explained, before the final result is arrived at in section~\ref{sec_monstarfr}. The theoretical result is then compared to results from computer simulations in section~\ref{sec_sims}. Finally, section~\ref{sec_concl} provides a summary and conclusions.

		\section{Gaussian fields}
	\label{sec_gsn_fields}

First, we will summarize the main definitions and characteristics of Gaussian fields that we need; see \cite{cite_paper1} for a more detailed description.

A homogeneous and isotropic Gaussian field can be defined as
\begin{equation}
  H(\vec{r})  =  \sum_{\vec{k}} A(k) \cos(\vec{k} \cdot \vec{r} + \phi_{\vec{k}}).
  \label{eq_gsn_field}
\end{equation}
The phases $\phi_{\vec{k}}$ are uniformly distributed random variables and independent of each other. The \emph{amplitude spectrum} $A(k)$ depends only on the length of the wave vector $\vec{k}$ and gives the Gaussian field its special characteristics, often expressed in terms of the \emph{moments} of the spectrum
\begin{equation}
  K_n  =  \sum_{\vec{k}} \tfrac12 A(k)^2 k^n.
\end{equation}
If the spectrum is sufficiently smooth -- which we will consider to always be the case here -- the sum can be replaced by an integral,
\begin{equation}
  K_n  =  \int \! \rd k \, \Pi(k) k^n.
\end{equation}
Here the integration over the polar coordinate of $\vec{k}$ has been absorbed into $\Pi(k)$, which is the \emph{power spectrum}. For convenience, we will consider $H$ to be normalized, such that $K_0 = \la H^2 \ra = 1$.

		\section{Umbilical points}
	\label{sec_umb}

Umbilical points are points on a surface where the curvature of the surface is the same along all directions. The curvature depicts how much the surface bends along a given direction, just like the second derivative of a one-dimensional function does. At an umbilical point then, the surface is locally spherical (or flat).

In order to make a proper mathematical formulation, consider a two-dimensional function $f(x,y)$. We consider any specific point $(x_0, y_0)$ and any direction given by an angle $\psi$. Along this direction, the function can be parametrized as
\begin{equation}
  f_\psi(r)  =  f(x_0 + r \cos\psi, y_0 + r \sin\psi).
\end{equation}
This function now describes what $f$ looks like at $(x_0, y_0)$ along the direction $\psi$. The \emph{curvature} is the value of the second derivative of $f_\psi(r)$ at $r = 0$,
\begin{IEEEeqnarray}{rLl}
  & \mc{2}{l}{ f_\psi''(0)  =  \frac{\rd^2 f_\psi}{\rd r^2} \Big|_{r=0} } \nl
  & \quad =	& f_{xx}(x_c, y_c) \cos^2 \psi + f_{yy}(x_c, y_c) \sin^2 \psi \nl
  &		& +\: 2 f_{xy}(x_c, y_c) \sin\psi \cos\psi \nl
  & \quad =	& (\tfrac12 + \tfrac12 \cos 2\psi) f_{xx} + (\tfrac12 - \tfrac12 \cos 2\psi) f_{yy} + \sin 2\psi f_{xy} \nl
  & \quad =	& \tfrac12 (f_{xx} + f_{yy}) + \tfrac12 (f_{xx} - f_{yy}) \cos 2\psi + f_{xy} \sin 2\psi.
  \label{eq_curv_orig}
\end{IEEEeqnarray}
We can write this in a more lucid form by applying the transformation
\begin{align}
  \tfrac12 (f_{xx} - f_{yy})	& = R \cos \alpha	& R		& = \tfrac12 \sqrt{(f_{xx}-f_{yy})^2 + 4 f_{xy}^2} \nl
  f_{xy}			& = R \sin \alpha	& \tan \alpha	& = \frac{2 f_{xy}}{f_{xx}-f_{yy}}
  \label{eq_curv_trf}
\end{align}
With this we find
\begin{equation}
  \begin{split}
    f_\psi''(0) =\:	& \tfrac12 (f_{xx} + f_{yy}) \\
    			& + \tfrac12 \sqrt{(f_{xx}-f_{yy})^2 + 4 f_{xy}^2} \cos(2\psi - \alpha).
  \end{split}
  \label{eq_curvature}
\end{equation}

With the curvature now properly defined, we introduce the two \emph{principal directions}, which are the directions along which the curvature is maximal or minimal. The corresponding curvatures are known as the \emph{principal curvatures}. We can easily see from eq.~\eqref{eq_curvature} that these two directions are given by $2\psi - \alpha = k\pi$ and hence perpendicular to each other.

As noted before, at an umbilical point the curvature is the same along all directions. In other words, the two principal curvatures are the same, and the principal directions cannot be defined. From eq.~\eqref{eq_curvature}, the definition of an umbilical point is easily seen to be
\begin{equation}
  f_{xx}  =  f_{yy}  \qquad \text{and} \qquad  f_{xy}  =  0.
  \label{eq_umbilic}
\end{equation}

Umbilical points can be classified in three types. The distinction can be clearly made when one looks at the \emph{curvature lines}. These are curves which are always tangent to a principal direction, either the one corresponding with the maximal curvature or the minimum one. These two sets of curvature lines intersect at right angles, since as noted before, the principal directions are always perpendicular to each other.

\begin{figure}
  \centering
  \subfigure[]{\includegraphics{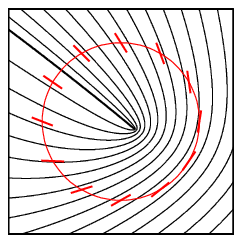}}
  \subfigure[]{\includegraphics{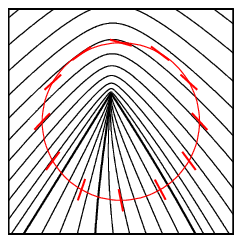}}
  \subfigure[]{\includegraphics{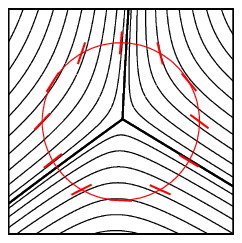}}
  \caption{One set of curvature lines (black) around (a) a lemon, (b) a monstar and (c) a star. The other set shows the same pattern in all cases. The red circle and line segments show how the principal direction rotates around the umbilical point.}
  \label{fig_umbilics}
\end{figure}

At an umbilical point, no principal direction can be defined, giving one of the three patterns shown in figure~\ref{fig_umbilics}. There are three types: \emph{lemons}, \emph{monstars} and \emph{stars}.

We see that, in each case, the umbilical point is a \emph{topological defect}, having a \emph{topological index} (see \cite{cite_Chaikin}, for example). Formally, the topological index is defined as
\begin{equation}
  n = \frac1{2\pi} \oint \! \nabla \psi \cdot \rd l,
  \label{eq_top_index}
\end{equation}
where the path-integral is taken over an (infinitesimal) counterclockwise loop around the defect. In words, it counts the number of revolutions the principal direction makes when traversing this closed loop. We see from fig.~\ref{fig_umbilics} that, for the point labeled \emph{star}, the direction makes half a clockwise rotation, which means a topological index of $-1/2$. The minus sign reflects that it rotates in the opposite direction with respect to the direction the loop is traversed in. The other two umbilical points have index $+1/2$.

Another characteristic separating the three is the number of curvature lines that terminate at the umbilical point. For a \emph{lemon}, this is one, whereas for the other two it is three. We see that the third type of umbilical point shares properties with both others: it has topological index $+1/2$, as does a lemon, and three curvature lines terminating at it, like a star. This in-between nature of the point is reflected in its name: \emph{monstar}.


The three types of umbilical point can also be distinguished using the third derivatives, much like the various types of critical points can be identified by the second derivatives. From eqs.~\eqref{eq_curv_trf} and \eqref{eq_curvature} we see that the two principal directions, for which the curvature is maximal or minimal, are given by
\begin{equation}
  \tan 2\psi = \frac{2f_{xy}}{f_{xx}-f_{yy}}.
  \label{eq_princ_dir}
\end{equation}
Note that the directions are given by angles modulo $\pi$, so this equation has two solutions: the two principal directions. The angle $2\psi$ can be pictured as the argument of the vector $\vec{v} = \tvec{f_{xx}-f_{yy}}{2f_{xy}}$. At an umbilical point, both vector components are zero (hence the angle / principal direction is not defined). In order to determine the topological index, we need to know what the principal directions are in close proximity to this point, in order to evaluate the infinitesimal loop in eq.~\eqref{eq_top_index}. We can expand $\vec{v}$ using the third derivatives. For a point $\vec{r}$ near an umbilical point $\vec{r_0}$ we have
\begin{equation}
  \vec{v}  =
  \begin{pmatrix}
    f_{xxx}-f_{yyx}	& f_{xxy}-f_{yyy} \\
    2f_{xyx}		& 2f_{xyy}
  \end{pmatrix}
  \begin{pmatrix}
    x-x_u \\
    y-y_u \\
  \end{pmatrix}
  =  \mathbf{A}(\vec{r} - \vec{r_0}).
\end{equation}

If $\mathbf{A}$ were the identity matrix, then a counterclockwise loop around $\vec{r_0}$ would obviously result in $2\psi$ increasing by $2\pi$, giving index $+1/2$. In general, $\mathbf{A}$ may shear and rotate $\vec{r}$, or may reflect it. The former would have no effect on the charge. However, if $\mathbf{A}$ includes a reflection, the gradient would rotate in the opposite direction and the index becomes $-1/2$. Whether $\mathbf{A}$ describes a reflection or not is encoded in the sign of its determinant,
\begin{equation}
  \tfrac12 \det \mathbf{A}  =  (f_{xxx}-f_{xyy})f_{xyy} + (f_{yyy}-f_{xxy})f_{xxy}.
\end{equation}
Hence, the index of the umbilical point is $+1/2$ ($-1/2$) if $\det \mathbf{A}$ is positive (negative). Introducing $\alpha = f_{xxx}, \beta = f_{xxy}, \gamma = f_{xyy}, \delta = f_{yyy}$, we thus find (see also \cite{cite_Berry})
\begin{equation}
  \alpha\gamma - \gamma^2 + \beta\delta - \beta^2
  \begin{cases}
    > 0 & \text{for L, M} \\
    < 0 & \text{for S}
  \end{cases}
  \label{eq_LM-S}
\end{equation}

As mentioned before, the criterion separating the lemons from the monstars (and stars), is the number of (locally straight) lines ending at the umbilical point: one for lemons, three for (mon)stars (that is one/three for each principal direction). This can also be expressed in terms of $\alpha$, $\beta$, $\gamma$ and $\delta$. Consider again a point $\vec{r}$ near $\vec{r_0}$. The principal directions are given by $\psi$ modulo $\tfrac12\pi$, hence one of the two is directed toward the umbilical point when the argument $\alpha$ of $\vec{r}-\vec{r_0}$ is equal to $\psi$ at $\vec{r}$, modulo $\tfrac12\pi$.

To find an algebraic statement of this condition, we double both sides: $2\alpha \equiv 2\psi \ (\mathrm{mod}\ \pi)$. The right-hand side is the argument of $\vec{v}$ by eq.~\eqref{eq_princ_dir}. We can also find a vector whose argument is given by the left-hand side: let $\tvec{x}{y} = \vec{r}-\vec{r_0}$; then $2\alpha$ is the argument of the vector $\tvec{x^2-y^2}{2xy}$ -- this is easily seen by mapping $\tvec{x}{y}$ to the complex number $x+iy$ and taking its square, which doubles the argument. The condition for $\vec{r}$ being on a terminating curvature line is that the arguments of these two vectors must match modulo $\pi$, which translates to $\tvec{x^2-y^2}{2xy}$ and $\mathbf{A}\tvec{x}{y}$ being parallel to each other. This condition can be mathematically expressed using the matrix $\mathbf{A}$ from before and the cross-product, giving
\begin{align}
  0	& = \tfrac12
  	    \begin{pmatrix}
  	      -2xy	& x^2-y^2
  	    \end{pmatrix}
  	    \mathbf{A}
  	    \begin{pmatrix}
  	      x \\
  	      y
  	    \end{pmatrix}
  	    \nl
  	& = \tfrac12
  	    \begin{pmatrix}
  	      -2xy	& x^2-y^2
  	    \end{pmatrix}
  	    \begin{pmatrix}
  	      (\alpha-\gamma)x - (\delta-\beta)y \\
  	      2\beta x + 2\gamma y
  	    \end{pmatrix}
  	    \nl
  	& = \beta x^3 - (\alpha-2\gamma) x^2y + (\delta-2\beta) xy^2 - \gamma y^3.
\end{align}
Note that this equation describes lines passing through $\vec{r_0}$, whereas the curvature lines actually terminate on the defect. On one side of $\vec{r_0}$, this line corresponds with the line of maximal curvature, on the other side with minimal curvature. This is easily seen from eq.~\eqref{eq_curv_orig}, if one notes that the second derivatives change sign when passing through $\vec{r_0}$.

The number of straight lines passing through $\vec{r_0}$ is thus equal to the number of (real) roots of this cubic equation (that is, by interpreting this as an equation in $x/y$). This is captured by the discriminant: if it is positive, then there are three roots; if it is negative, there is only one. This results in (see also \cite{cite_Berry})
\begin{equation}
  \renewcommand{\arraystretch}{1.2}
  \begin{array}{l}
    4 \big( 3\gamma(\alpha-2\gamma)-(\delta-2\beta)^2 \big) \\
    \times \big( 3\beta(\delta-2\beta)-(\alpha-2\gamma)^2 \big) \\
    - \: \big( (\delta-2\beta)(\alpha-2\gamma)-9\beta\gamma \big) ^2
  \end{array}
  \begin{cases}
    > 0 & \text{for M, S} \\
    < 0 & \text{for L}
  \end{cases}
  \label{eq_MS-L}
\end{equation}
In summary:

\vspace{\baselineskip}
{ \renewcommand{\arraystretch}{1.5}
  \newcommand{\cc}{\centering}
  \newcommand{\tnl}{\tabularnewline}
  \begin{tabular}{|p{2.5cm}|p{2.5cm}|p{2.5cm}|}
    \hline
    \multicolumn{3}{|c|}{ Umbilical point }							\tnl
    \hline
    \multicolumn{2}{|c|}{ Index $+\tfrac12$ }		& \cc Index $-\tfrac12$			\tnl
    \hline
    \cc Lemon				& \cc Monstar	& \cc Star				\tnl
    \hline
    \multicolumn{2}{|c|}{ eq.~\eqref{eq_LM-S} $ > 0$ }	& \cc eq.~\eqref{eq_LM-S} $ < 0$	\tnl
    \hline
    \cc eq.~\eqref{eq_MS-L} $ < 0$	& \multicolumn{2}{c|}{ eq.~\eqref{eq_MS-L} $> 0$ }	\tnl
    \hline
  \end{tabular}
}
\vspace{\baselineskip}

According to the Poincar\'e-Hopf theorem, for any surface the total sum of all topological indices equals the Euler characteristic of the underlying manifold. For the two-dimensional plane that we consider, this is simply zero. As a consequence, the density of stars (with index $+1/2$) equals the combined density of lemons and monstars (with index $-1/2$). In other words, the star \emph{fraction}, that is the density of stars divided by the total density of umbilical points, is always $1/2$. There is however no topological constraint on the lemon and monstar fractions. For \emph{isotropic} Gaussian random fields however, it has been shown that the monstar fraction is $\alpha_M = 1/2 - 1/\sqrt{5} = 0.053$ \cite{cite_Berry, cite_Dennis2}, independent of the spectrum of the Gaussian field. As a consequence, the monstar fraction promises to be a good criterion to test the (non-)Gaussianity of a field. Should one be given a random field, and find that the monstar fraction is not equal to $0.053$, one can immediately conclude that the field is not an isotropic Gaussian one.

In the next sections we take a \emph{non}-Gaussian field $h$ that can be described as a Gaussian field $H$ with a perturbation $f(H)$ added to it, and calculate how much the monstar fraction $\alpha_M$ deviates from the universal value $0.053$ as a function of the perturbation $f(H)$. Our result also allows to attack the reverse problem: when given a non-Gaussian field of which the type of perturbation is known, we can determine the monstar fraction and thus reveal the size of the perturbation.

		\section{The generating function}
	\label{sec_gen_func}

We consider a field of the form $h(\vec{r}) = H(\vec{r}) + f(H(\vec{r}))$, where $H(\vec{r})$ is a Gaussian field and $f$ a small nonlinear function of $H(\vec{r})$ only. As we have seen in eqs.~\eqref{eq_umbilic}, \eqref{eq_LM-S} and \eqref{eq_MS-L}, the monstars can be defined using the second and third derivatives of the field $h$ with respect to $x$ and $y$. Determining the monstar fraction thus boils down to determining how likely it is that at a specific point $\vec{r}$ the third derivatives $\alpha = h_{xxx}(\vec{r})$, $\beta = h_{xxy}(\vec{r})$, $\gamma = h_{xyy}(\vec{r})$ and $\delta = h_{yyy}(\vec{r})$ are such that eqs.~\eqref{eq_LM-S} and \eqref{eq_MS-L} prescribe a monstar, given that the second derivatives obey $h_{xx}(\vec{r}) = h_{yy}(\vec{r})$ and $h_{xy}(\vec{r}) = 0$.

In order to determine this, we require the joint probability distribution of these seven stochastic variables. When we have this, we can set $h_{xx} = h_{yy}$ and $h_{xy} = 0$ and integrate $\alpha$, $\beta$, $\gamma$ and $\delta$ over the appropriate ranges to get the density of monstars and all umbilical points respectively. The ratio of these then gives the monstar fraction.

We shall arrive at the desired probability distribution by determining the corresponding generating function, which is defined as the Fourier transform of the probability distribution \cite{cite_Kampen}. For a set of $n$ correlated variables $\{h_i\}$ this is
\begin{IEEEeqnarray}{rLl}
  & \mc{2}{l}{ \chi(\lambda_1, \ldots, \lambda_n) } \nl
  & \qquad =	& \int \! \rd h_1 \ldots \rd h_n \, p(h_1, \ldots, h_n) e^{i(h_1\lambda_1 + \ldots + h_n\lambda_n)} \nl
  & \qquad =	& 1 + i \sum_j \la h_j \ra \lambda_j + \frac{i^2}{2!} \sum_{j_1,j_2} \la h_{j_1}h_{j_2} \ra \lambda_{j_1}\lambda_{j_2} \nl
  &		& +\: \frac{i^3}{3!} \sum_{j_1,j_2,j_3} \la h_{j_1}h_{j_2}h_{j_3} \ra \lambda_{j_1}\lambda_{j_2}\lambda_{j_3} + \ldots
  \label{eq_gen_func}
\end{IEEEeqnarray}
Here, the coefficients $\la \ldots \ra$ are the \emph{moments}, or multivariable correlations defined by
\begin{equation}
  \la h_{j_1} \ldots h_{j_k} \ra	 \equiv \int \! \rd h_1 \ldots \rd h_n \, p(h_1, \ldots, h_n) h_{j_1} \ldots h_{j_k}.
\end{equation}
Eq.~(\ref{eq_gen_func}) is proved by expanding the exponential term by term.

Upon taking the logarithm of $\chi$ and expanding, the quantities known as the \emph{cumulants} are revealed:
\begin{align}
  \log \chi =\:	& i \sum_j C_1(h_j)\lambda_j + \frac{i^2}{2!} \sum_{j_1,j_2} C_2(h_{j_1},h_{j_2})\lambda_{j_1}\lambda_{j_2} \nl
  		& + \frac{i^3}{3!} \sum_{j_1,j_2,j_3} C_3(h_{j_1},h_{j_2},h_{j_3})\lambda_{j_1}\lambda_{j_2}\lambda_{j_3} + \ldots
  \label{eq_log_chi}
\end{align}
The cumulants can be written in terms of the moments, as can be seen by taking the logarithm of eq.~\eqref{eq_gen_func} and expanding it. For example,
\begin{equation}
  \begin{split}
    C(h_1, h_2, h_3) =\:	& \la h_1h_2h_3 \ra - \la h_1 \ra \la h_2h_3 \ra - \la h_2 \ra \la h_3h_1 \ra \\
    				& - \la h_3 \ra \la h_1h_2 \ra + 2 \la h_1 \ra \la h_2 \ra \la h_3 \ra.
  \end{split}
  \label{eq_cum_example}
\end{equation}
In reverse, the moments can be written in terms of the cumulants, e.g.
\begin{equation}
  \begin{split}
    \la h_1h_2h_3 \ra =\:	& C(h_1, h_2, h_3) + C(h_1) C(h_2, h_3) \\
    				& + C(h_2) C(h_3, h_1) + C(h_3) C(h_1, h_2) \\
    				& + C(h_1) C(h_2) C(h_3).
  \end{split}
  \label{eq_sum_cumulants}
\end{equation}
If all the moments or all the cumulants are known, we can construct the generating function and perform an inverse Fourier transformation to obtain the probability distribution.

The defining characteristic of Gaussian random variables $H_i$ is that all cumulants are zero, with the exception of the second-order ones $C_2(H_i, H_j) = \la H_i H_j \ra$. In this case, the generating function is thus
\begin{equation}
  \chi(\lambda_1, \ldots, \lambda_n)  =  \exp \Big( \!-\! \tfrac12 \sum_{ij} C_2(H_i, H_j) \lambda_i \lambda_j \Big).
  \label{eq_chi_gauss}
\end{equation}
The inverse Fourier transformation yields the standard distribution for correlated Gaussian random variables (see e.g.\ \cite{cite_Longuet1}),
\begin{equation}
  \begin{split}
    p(H_1, \ldots, H_n) =\:	& \frac1{(2\pi)^{n/2} \sqrt{\det{\sigma}}} \\
    				& \times \exp \Big( \!-\! \tfrac12 \sum_{i,j}{(\sigma^{-1})_{ij} H_i H_j} \Big),
  \end{split}
  \label{eq_gauss_distr}
\end{equation}
where $\sigma$ is the matrix of correlations, $\sigma_{ij}  =  \la H_i H_j \ra$.

For a Gaussian field $H$, the derivatives are themselves Gaussian fields -- therefore the above formula gives their joint distribution.
For the non-Gaussian field $h$, there are some small corrections to this distribution. To find these corrections to first order, we need to determine the cumulants to first order in $f(H)$. We will see that only a small number of cumulants are nonzero up to this order. Before we proceed to derive them, we switch to a complex coordinate system which allows for optimal usage of translational and rotational symmetry, which $h$ has inherited from $H$ for the type of perturbations under consideration.

		\section{Complex coordinates representation}
	\label{sec_complex}

To find the distribution of umbilical points, we now have to find the joint distribution of the seven second and third derivatives of $h$. All these variables can be combined into a more compact form by using complex coordinates.
These will make it easier to evaluate the integral that determines the monstar density, and will help us to work out the probability distribution with the help of symmetry.

The complex coordinates are given by
\begin{align}
  z	& =  x + iy	& x	& =  \tfrac12(z+z^*) \nl
  z^*	& =  x - iy	& y	& =  \tfrac12i(z^*-z)
\end{align}
Of course, as complex numbers, $z$ and $z^*$ are not independent; however we can formally define partial derivatives with respect to each of them, using the chain rule, just like we could if this transformation involved a real number instead of $i$.

The derivatives with respect to $z$ and $z^*$ are given by
\begin{align}
  \frac{\partial}{\partial z}	& = \tfrac12 \frac{\partial}{\partial x} - \tfrac12i \frac{\partial}{\partial y},	& \frac{\partial}{\partial z^*}	& = \tfrac12 \frac{\partial}{\partial x} + \tfrac12i \frac{\partial}{\partial y}.
\end{align}
We see that the derivatives with respect to $z$ and $z^*$ are each other's conjugate, but again, we consider both to be linear transformations of $\partial_x$ and $\partial_y$.

The usefulness of using $z$ and $z^*$ can be immediately seen from $h_{zz}$:
\begin{equation}
  h_{zz}  =  \partial_z^2 h  =  \tfrac14 (\partial_x - i\partial_y)^2 h  =  \tfrac14 (h_{xx} - h_{yy} + 2i h_{xy}).
\end{equation}
We see that the definition of an umbilical point can be captured in one equation: $h_{zz} = 0$.

The various types of umbilical points were defined in eqs.~\eqref{eq_LM-S} and \eqref{eq_MS-L} using the ``normal'' third derivatives $h_{xxx} = \alpha$, $h_{xxy} = \beta$, $h_{xyy} = \gamma$ and $h_{yyy} = \delta$. In terms of $h_{zzz}$, $h_{zzz^*}$, $h_{zz^*z^*}$ and $h_{z^*z^*z^*}$ the two conditions for a monstar become
\begin{subequations}
  \begin{align}
    |h_{zzz^*}|^2 - |h_{zzz}|^2						& > 0, \\
    27|h_{zzz}|^4 - |h_{zzz^*}|^4 - 18|h_{zzz}|^2 |h_{zzz^*}|^2 \qquad	& \nl
    -\: 4(h_{zzz}h_{zz^*z^*}^3 + h_{z^*z^*z^*}h_{zzz^*}^3)		& > 0.
  \end{align}
  \label{eq_monstar}
\end{subequations}
Here $|h_{zzz}|^2$ and $|h_{zzz^*}|^2$ represent $h_{zzz}h_{z^*z^*z^*}$ and $h_{zzz^*}h_{zz^*z^*}$ respectively. 

The density of monstars will essentially be given by integrating the probability distribution $p(h_{zz}=0,h_{zzz},h_{zzz^*})$ over the range defined by these conditions. This probability distribution is determined by the cumulants of combinations of the three variables. Rotational and translational symmetry however imply that only a small number of these combinations yield a nonzero cumulant.

First, consider the consequences of the isotropy (rotational symmetry) of the field $h(\vec{r})$ for a moment like $\la h_{z^*}(\vec{r})h_{zz}(\vec{r}) \ra$. Note that, due to homogeneity (translational symmetry), this moment does not depend on $r$; it will often be dropped from now on. 
Isotropy implies that this moment should not change if we rotate the field around $\vec{r}$, over any angle $\alpha$. In terms of $z$ and $z^*$, this results in the transformation
\begin{align}
  z'	& =  e^{i\alpha}z	& \partial_{z'}		& =  e^{-i\alpha}\partial_z \nl
  z'^*	& =  e^{-i\alpha}z^*	& \partial_{z'^*}	& =  e^{i\alpha}\partial_{z^*}
  \label{eq_rotation}
\end{align}
As a result, we get $\la h_{z'}h_{z'^*z'^*} \ra = e^{i\alpha} \la h_zh_{z^*z^*} \ra$. Since we argued that the two expectation values must be equal, for any $\alpha$, we must have $\la h_{z^*}h_{zz} \ra = 0$.

In general, following a rotation expectation values pick up a factor $e^{ik\alpha}$, where $k$ is the number of $z^*$ derivatives inside the bracket minus the number of $z$ derivatives. By the above argument, the expectation value is zero if $k \neq 0$. Therefore, an expectation value can only be nonzero if the numbers of $z$ and $z^*$ derivatives inside the bracket are equal. Since a cumulant is a sum of products of expectation values, featuring every variable once in every product (compare eq.~\eqref{eq_cum_example}), the same property applies to cumulants.

The homogeneity (translational symmetry) of the fields under consideration provides another useful trick that relates different cumulants to one another. As already stated, a moment like $\la h_1(\vec{r}) \ldots h_n(\vec{r}) \ra$ does not depend on $\vec{r}$. Hence the derivative of this with respect to $z$ or $z^*$ is zero. Applying the product rule:
\begin{align}
  0	& = \partial_z \la h_1 \ldots h_n \ra \nl
  	& = \la (\partial_z h_1) h_2 \ldots h_n \ra + \ldots + \la h_1 h_2 \ldots (\partial_z h_n) \ra.
\end{align}
For $n=2$, this gives the useful relation
\begin{equation}
  \la (\partial_z h_1) h_2 \ra  =  - \la h_1 (\partial_z h_2) \ra.
\end{equation}
In essence, for a two-point correlation it is possible to ``transfer'' a $z$ derivative from the one term to the other at the cost of an overall minus sign. The same applies of course to a $z^*$ derivative. For example, we find the relation $\la h_{zz}h_{z^*z^*} \ra  =  - \la h_zh_{zz^*z^*} \ra$.

Together, these two symmetries constrain the probability distribution $p(h_{zz},h_{zzz},h_{zzz^*})$. In particular, they explain why the monstar fraction is always the same for any Gaussian distribution \cite{cite_Berry}. A Gaussian distribution does not have many degrees of freedom to start with; only the two-point correlations between the variables are adjustable. In this case, the two-point correlations between any two of these variables is zero, by rotational symmetry, while the variances of $h_{zzz}$ and $h_{zzz^*}$ are equal by translational symmetry. Hence (after setting $h_{zz}=0$ to identify the umbilical points), the distribution $p(h_{zz}=0,h_{zzz},h_{zzz^*})$ is always the same apart from a scale, and that determines the monstar fraction. This argument can be generalized to singularities in the polarization field of light (even though the field might not be derived from a scalar field $h$), and so Gaussian polarization fields have the same monstar fraction as well, as shown in \cite{cite_Dennis2}.

On the other hand, when $h$ has non-Gaussian contributions, there are many more cumulants, and symmetry is not enough to constrain them any more. For the field $h=H+f(H)$ we are studying, we proceed to calculate the cumulants explicitly.

		\section{The cumulants}
	\label{sec_cumulants}

Although the problem is now cast in terms of complex derivatives, the recipe outlined in section~\ref{sec_gen_func} still applies. The task is to determine the cumulants of $h_{zz}$, $h_{zzz}$, $h_{zzz^*}$ and their conjugates up to first order in the perturbation $f$.

These cumulants have the general form $C_n(D_1 h, \ldots, D_n h)$, where each $D_j$ represents a number of $z$ and $z^*$ derivatives. For the moment, let us consider each $D_j h$ to be at a different point $\vec{r_j}$, i.e.\ $C_n(D_1 h(\vec{r_1}), \ldots, D_n h(\vec{r_n}))$. Later we will set all points equal again. For convenience, we shall drop the vector notation, i.e.\ $r_i = \vec{r_i}$. Since now each derivative $D_j$ acts only at a specific point, we can bring them outside the cumulant:
\begin{equation}
  \begin{split}
    & C_n(D_1 h, \ldots, D_n h) \\
    & \qquad = D_1 \ldots D_n C_n(h(r_1), \ldots, h(r_n)) \Big|_{r_1 = \ldots = r_n}.
  \end{split}
  \label{eq_cum_deriv}
\end{equation}

Let us write $h(r_j) = h_j$ for shortness, and focus on $C_n(h_1, \ldots, h_n)$. Inserting $h_j = H_j + f(H_j)$, expanding the cumulant and keeping only terms up to first order in $f$ yields
\begin{equation}
  \begin{split}
    C_n(h_1, \ldots, h_n) =\:	& C_n(H_1, \ldots, H_n) \\
    				& + C_n(f(H_1), H_2, \dots, H_n) \\
    				& + C_n(H_1, f(H_2), \ldots, H_n) + \ldots \\
    				& + C_n(H_1, H_2, \ldots, f(H_n)).
  \end{split}
  \label{eq_cumulant_sum}
\end{equation}

The first term on the right-hand side is now simply the cumulant of a set of Gaussian random variables, which, as discussed before, is zero for $n>2$. The other terms can be evaluated perturbatively and are equivalent to each other. Consider the second term as an example. For a cumulant involving Gaussian variables and one function of a Gaussian we have (see Appendix~\ref{app_cum_with_f}):
\begin{equation}
  \begin{split}
    & C_n(f(H_1), H_2, \ldots, H_n) \\
    & \qquad = \la f^{(n-1)}(H_1) \ra \la H_1H_2 \ra \la H_1H_3 \ra \ldots \la H_1H_n \ra.
  \end{split}
  \label{eq_cum_with_f}
\end{equation}
When we reinsert the derivatives $D_2$ through $D_n$ from eq.~\eqref{eq_cum_deriv} and set $r_2 = \ldots = r_n = r$ we get
\begin{equation}
  \begin{split}
    & D_1 \ldots D_n C_n(f(H(r_1)), \ldots, H(r_n)) \Big|_{r_1 = \ldots = r_n} \\
    & \qquad  =  D_1 \la f^{(n-1)}(H_1) \ra \la H_1 D_2H \ra \ldots \la H_1D_nH \ra \Big|_{r_1=r}.
  \end{split}
  \label{eq_cumulant_gsn_func}
\end{equation}

Now we can reinsert $D_1$ and then set $r_1=r$, as prescribed by eq.~\eqref{eq_cum_deriv}. Remember that $D_1$ only acts on $H_1$. Due to the product rule, we have to consider all possible ways in which the derivatives in $D_1$ can be distributed over all $H_1$'s. Recall from section \ref{sec_complex} that, after setting $r_1=r$, each expectation value can only be nonzero if the number of $z$ and $z^*$ derivatives inside are equal. Note also that $\la f^{(n-1)}(H_1) \ra$ does not depend on $r_1$, hence any derivative of it is zero. Therefore, the only nonzero contributions stemming from the product rule are those distributions that make the number of $z$ and $z^*$ derivatives equal inside each bracket.

We consider the cumulant $C_3(h_{zz}, h_{zz^*z^*}, h_{zz^*z^*})$ as an example to demonstrate the procedure. First, we find
\begin{IEEEeqnarray}{rLl}
  & \mc{2}{l}{ C_3(h_{zz}, h_{zz^*z^*}, h_{zz^*z^*}) } \nl
  & \qquad =	& C_3(H_{zz}, H_{zz^*z^*}, H_{zz^*z^*}) \nl
  &		& +\: \partial_{z^{}_1z^{}_1}(\la f''(H) \ra \la H_1H_{zz^*z^*} \ra \la H_1H_{zz^*z^*} \ra) \\
  &		& +\: 2 \partial_{z^{}_2z^*_2z^*_2} (\la f''(H) \ra \la H_2H_{zz} \ra \la H_2H_{zz^*z^*} \ra). \nonumber
  \label{cum_example}
\end{IEEEeqnarray}
The first term is zero, since all cumulants of Gaussian variables are zero beyond second order.
For the second term, we need to consider how to distribute the two $\partial_{z_1}$ derivatives to make all expectation values nonzero. The only possibility is to put one $\partial_{z_1}$ in front of each $H_1$. Note however that this term appears twice in the product rule, because there are two ways of distributing the two derivatives. After setting $r_1 = r$ we thus have
\begin{equation}
  \begin{split}
    & \partial_{z^{}_1z^{}_1}(\la f''(H) \ra \la H_1H_{zz^*z^*} \ra \la H_1H_{zz^*z^*} \ra) \\
    & \qquad = 2 \la f''(H) \ra \la H_zH_{zz^*z^*} \ra ^2.
  \end{split}
\end{equation}
In the third term on the right-hand side of eq.~\eqref{cum_example} we have one $\partial_z$ and two $\partial_{z^*}$'s to distribute. The first $H_2$ needs $\partial_{z^*z^*}$ to balance the derivatives and the other takes the $\partial_z$ derivative. There are no multiple ways to distribute these derivatives in this case and therefore
\begin{align}
  & \partial_{z^{}_2z^*_2z^*_2} (\la f''(H) \ra \la H_2H_{zz} \ra \la H_2H_{zz^*z^*} \ra) \nl
  & \qquad = \la f''(H) \ra \la H_{z^*z^*}H_{zz} \ra \la H_zH_{zz^*z^*} \ra.
\end{align}
Combining everything together results in
\begin{align}
  & C_3(h_{zz}, h_{zz^*z^*}, h_{zz^*z^*}) \nl
  & \quad = 2 \la f''(H) \ra \la H_zH_{zz^*z^*} \ra (\la H_zH_{zz^*z^*} \ra + \la H_{z^*z^*}H_{zz} \ra).
\end{align}
Finally, due to translational symmetry, we have $\la H_zH_{zz^*z^*} \ra + \la H_{z^*z^*}H_{zz} \ra = \partial_z \la H_zH_{z^*z^*} \ra = 0$. We thus find that $C_3(h_{zz}, h_{zz^*z^*}, h_{zz^*z^*}) = 0$.

Now we will show that there are only a finite number of nonzero cumulants (up to first order in $f$). In fact, there are none beyond the fourth order. Consider eq.~\eqref{eq_cumulant_sum} with $n > 4$. The first term (zero order) is zero because it is the cumulant of more than two Gaussian variables. For the other ones we apply the recipe of eq.~\eqref{eq_cumulant_gsn_func}. We have $n-1$ brackets in which the $z$ and $z^*$ derivatives need to be matched. Since we are only considering the variables $h_{zz}$, $h_{zzz}$, $h_{zzz^*}$ and their conjugates, each one has a mismatch to begin with. However, since $D_1$ has only three derivatives at most, it is not possible to balance the derivatives in all $n-1$ brackets.

This ``lack of derivatives'' also kills a lot of cumulants of lower order, especially fourth order. For example, in $C_4(h_{zz}, h_{z^*z^*}, h_{zzz^*}, h_{zz^*z^*})$ the first two variables require two derivatives to balance the derivatives and the other two require one. Therefore, no matter from which variable the derivatives are distributed, there is always a shortage.

All the nonzero cumulants are listed in table~\ref{tbl_cumulants}. Two parameters were introduced:
\begin{subequations}
  \begin{align}
    \sigma	& \equiv \la H_{zz}H_{z^*z^*} \ra  =  -\la H_zH_{zz^*z^*} \ra, \\
    \tau	& \equiv \la H_{zzz}H_{z^*z^*z^*} \ra  =  \la H_{zzz^*}H_{zz^*z^*} \ra.
  \end{align}
\end{subequations}
Note that the trick based on translational symmetry was used to equate the expectation values. In the second equation, it was used twice (transferring a $\partial_z$ one way and a $\partial_{z^*}$ the other way).

\begin{table}
  \centering
  \begin{tabular}{l @{\ \ =\ \ } l}
    $C_2(h_{zz}, h_{z^*z^*})$						&	$\sigma(1 + 2 \la f'(H) \ra)$	\\
    $C_2(h_{zzz}, h_{z^*z^*z^*})$					&	$\tau(1 + 2 \la f'(H) \ra)$	\\
    $C_2(h_{zzz^*}, h_{zz^*z^*})$					&	$\tau(1 + 2 \la f'(H) \ra)$	\\
    $C_3(h_{zz}, h_{zzz^*}, h_{z^*z^*z^*})$ + conj.			&	$-3 \sigma^2 \la f''(H) \ra$	\\
    $C_4(h_{zzz^*}, h_{zzz^*}, h_{zz^*z^*}, h_{zz^*z^*})$		&	$-8 \sigma^3 \la f'''(H) \ra$	\\
    $C_4(h_{zzz}, h_{zz^*z^*}, h_{zz^*z^*}, h_{zz^*z^*})$ + conj.	&	$-6 \sigma^3 \la f'''(H) \ra$	\\
  \end{tabular}
  \label{tbl_cumulants}
  \caption{All nonzero cumulants. The two asymmetric cumulants have a conjugate twin in which all $z$'s and $z^*$'s are interchanged.}
\end{table}

The parameters $\sigma$ and $\tau$ are related to the moments $K_n$ of $H$. This is most easily accomplished by writing $H$ in complex variables:
\begin{align}
  H	& = \sum_{\vec{k}} A(k) \cos(\vec{k} \cdot \vec{r} + \phi_{\vec{k}}) \nl
  	& = \sum_k A(|k|) \cos\big( \tfrac12 (k^*z + kz^*) + \phi_k \big).
\end{align}
Here $k = k_x + i k_y$ is the complex analogue of $\vec{k}= \tvec{k_x}{k_y}$. With this we find
\begin{subequations}
  \begin{align}
    H_{zzz}		& = \sum_k A(k) \tfrac18 (k^*)^3 \sin\big( \tfrac12 (k^*z + kz^*) + \phi_k \big), \\
    H_{z^*z^*z^*}	& = \sum_k A(k) \tfrac18 k^3 \sin\big( \tfrac12 (k^*z + kz^*) + \phi_k \big).
  \end{align}
\end{subequations}
Hence:
\begin{equation}
  \tau  =  \la H_{zzz}H_{z^*z^*z^*} \ra  =  \la \tfrac18 (k^*)^3 \tfrac18 k^3 \ra  =  \tfrac1{64}K_6.
\end{equation}
Similarly, we have $\sigma = \frac1{16}K_4$.

		\section{Probability distribution}
	\label{sec_prob_distr}

With the aid of the cumulants we can build the logarithm of the generating function (see eq.~\eqref{eq_log_chi}), provided that we identify the appropriate variables in Fourier space. Consider $h_{zz}$ and $h_{z^*z^*}$ for example. These complex variables represent two real variables $\xi_x$ and $\xi_y$, the real and imaginary part of $h_{zz}$. Let $\lambda_x$ and $\lambda_y$ be their Fourier counterparts. The generating function is
\begin{align}
  \chi(\lambda_x, \lambda_y, \ldots)	& = \int \! \rd \xi_x \rd \xi_y \ldots \, p(\xi_x, \xi_y, \ldots) e^{i(\xi_x\lambda_x + \xi_y\lambda_y + \ldots)} \nl
  					& = \la e^{i(\xi_x\lambda_x + \xi_y\lambda_y + \ldots)} \ra.
\end{align}
The exponent can be written in terms of the complex variables,
\begin{equation}
  \xi_x\lambda_x + \xi_y\lambda_y  =  h_{zz} \lambda_{zz}^* + h_{z^*z^*} \lambda_{zz},
  \label{eq_lambda}
\end{equation}
where we define $\lambda_{zz}=\frac{1}{2}(\lambda_x+i\lambda_y)$. Then $\lambda_{zz}$ is the complex Fourier variable corresponding to $h_{z^*z^*}$ and we likewise introduce $\lambda_{zzz}$ and $\lambda_{zzz^*}$, which are conjugate to $h_{z^*z^*z^*}$ and $h_{zz^*z^*}$. We will define integrals with respect to the complex Fourier variables, e.g.\ with respect to $d^2h_{zz}$, as integrals over the real and imaginary parts of $h_{zz}$, and the inverse Fourier transform will be performed by integrating over the real and imaginary parts of the $\lambda$'s.

The generating function is thus
\begin{equation}
  \begin{split}
    \log \chi =	& -\: C_2(h_{zz}, h_{z^*z^*}) \lambda_{z^*z^*}\lambda_{zz} \\
    		& -\: C_2(h_{zzz}, h_{z^*z^*z^*}) \lambda_{z^*z^*z^*} \lambda_{zzz} \\
    		& -\: C_2(h_{zzz^*}, h_{zz^*z^*}) \lambda_{zz^*z^*} \lambda_{zzz^*} \\
    		& -\: i C_3(h_{zz}, h_{zzz^*}, h_{z^*z^*z^*}) \lambda_{z^*z^*} \lambda_{zz^*z^*} \lambda_{zzz} \\
    		& -\: i C_3(h_{zz}, h_{zzz^*}, h_{z^*z^*z^*}) \lambda_{zz} \lambda_{zzz^*} \lambda_{z^*z^*z^*} \\
    		& +\: \tfrac14 C_4(h_{zzz^*}, h_{zzz^*}, h_{zz^*z^*}, h_{zz^*z^*}) \lambda_{zz^*z^*}^2 \lambda_{zzz^*}^2 \\
    		& +\: \tfrac16 C_4(h_{zzz}, h_{zz^*z^*}, h_{zz^*z^*}, h_{zz^*z^*}) \lambda_{z^*z^*z^*} \lambda_{zzz^*}^3 \\
    		& +\: \tfrac16 C_4(h_{zzz}, h_{zz^*z^*}, h_{zz^*z^*}, h_{zz^*z^*}) \lambda_{zzz} \lambda_{zz^*z^*}^3.
  \end{split}
\end{equation}
Upon entering the cumulants from table~\ref{tbl_cumulants}:
\begin{IEEEeqnarray}{Rlll}
  \log \chi =	& \mc{3}{l}{ -\tilde{\sigma} \lambda_{zz} \lambda_{z^*z^*} - \tilde{\tau} (\lambda_{zzz} \lambda_{z^*z^*z^*} + \lambda_{zzz^*} \lambda_{zz^*z^*}) } \nl
  		& \mc{2}{l}{ +\: 3i \sigma^2 \la f''(H) \ra (\!\! }	& \lambda_{zz} \lambda_{zzz^*} \lambda_{z^*z^*z^*} \nl
  		&				&			& + \lambda_{z^*z^*} \lambda_{zz^*z^*} \lambda_{zzz}) \\
  		& -\: \sigma^3 \la f'''(H) \ra (	& \mc{2}{l}{ 2 \lambda_{zzz^*}^2 \lambda_{zz^*z^*}^2 + \lambda_{zzz} \lambda_{zz^*z^*}^3 } \nl
  		&				& \mc{2}{l}{ + \lambda_{z^*z^*z^*} \lambda_{zzz^*}^3). } \nonumber
\end{IEEEeqnarray}
Here $\tilde{\sigma} = \sigma(1+2\la f'(H) \ra)$ and $\tilde{\tau} = \tau(1+2\la f'(H) \ra)$ have been introduced. The factors in front of the cumulants are the factor $i^k/k!$ in eq.~\eqref{eq_log_chi} multiplied with the number of permutations of the $\lambda$'s.

To obtain the probability distribution, we take the exponential and perform the inverse Fourier transformation (see eq.~\eqref{eq_gen_func}). This gives an integral over the exponential of a polynomial of degree 4. However, all terms of degree three and four are of order $f$, so we can expand the exponential and be left with only square terms in the exponent. The result is
\begin{IEEEeqnarray}{Rllll}
  \chi =	& \mc{4}{l}{ \exp\big( \!-\! \tilde{\sigma} \lambda_{zz} \lambda_{z^*z^*} - \tilde{\tau} (\lambda_{zzz} \lambda_{z^*z^*z^*} + \lambda_{zzz^*} \lambda_{zz^*z^*}) \big) } \nl
  		& \times \Big(	& \mc{2}{l}{ 1 + 3i \sigma^2 \la f''(H) \ra (\!\! }	& \lambda_{zz} \lambda_{zzz^*} \lambda_{z^*z^*z^*} \nl
		&		&					&		& + \lambda_{z^*z^*} \lambda_{zz^*z^*} \lambda_{zzz}) \nl
  		& 		& -\: \sigma^3 \la f'''(H) \ra (	& \mc{2}{l}{ 2 \lambda_{zzz^*}^2 \lambda_{zz^*z^*}^2 + \lambda_{zzz} \lambda_{zz^*z^*}^3 } \nl
		&		&					& \mc{2}{l}{ + \lambda_{z^*z^*z^*} \lambda_{zzz^*}^3) \Big). }
\end{IEEEeqnarray}

Now we can take the inverse Fourier transform. Note that $\lambda_{zz}$ and $\lambda_{z^*z^*}$ are each other's conjugate. Upon integrating the real and imaginary parts of $\lambda_{zz}$ and imposing $\lambda_{z^*z^*} = \lambda_{zz}^*$ (the same procedure applies to the other two pairs of $\lambda$'s), one obtains
\begin{IEEEeqnarray}{rll}
  & \mc{2}{l}{ p(h_{zz}, h_{z^*z^*}, h_{zzz}, h_{z^*z^*z^*}, h_{zzz^*}, h_{zz^*z^*}) } \nl
  & \qquad = \int \!	& \frac{\rd^2 \lambda_{zz} \rd^2 \lambda_{zzz} \rd^2 \lambda_{zzz^*}}{\pi^6} \, \chi \nl
  &			& \times e^{-i(\lambda_{zz}h_{zz} + \lambda_{zzz}h_{zzz} + \lambda_{zzz^*}h_{zzz^*} + \mathrm{conj.})}.
\end{IEEEeqnarray}
Note that the denominator is $\pi^6$ rather than $(2\pi)^6$ because of the factor of $\frac{1}{2}$ in the definitions of the $\lambda$'s (see eq.~\eqref{eq_lambda}).

The Fourier transform of a Gaussian function multiplied with a polynomial is easy to perform by noting that multiplying by $\lambda$ in Fourier space is equivalent to taking a derivative in normal space:
\begin{equation}
  \int \! \rd \lambda \, \lambda^n f(\lambda) e^{-i \lambda h}  =  \bigg( i \frac{\partial}{\partial h} \bigg)^n \! \int \! \rd \lambda \, f(\lambda) e^{-i \lambda h}.
\end{equation}
The inverse Fourier transform of the Gaussian part of the generating function is
\begin{equation}
  \begin{split}
    & \int \! \frac{\rd^2 \lambda_{zz} \rd^2 \lambda_{zzz} \rd^2 \lambda_{zzz^*}}{\pi^6} \, e^{-\tilde{\sigma} |\lambda_{zz}|^2 - \tilde{\tau} (|\lambda_{zzz}|^2 + |\lambda_{zzz^*}|^2)} e^{-i(\ldots)} \\
    & \qquad  =  \frac1{\pi^3 \tilde{\sigma} \tilde{\tau}^2} \exp\Big( \!-\! \frac1{\tilde{\sigma}}|h_{zz}|^2 - \frac1{\tilde{\tau}}(|h_{zzz}|^2 + |h_{zzz^*}|^2) \Big),
  \end{split}
\end{equation}
and the final result reads
\begin{IEEEeqnarray}{Rllll}
  & \mc{4}{l}{ p(h_{zz}, h_{z^*z^*}, h_{zzz}, h_{z^*z^*z^*}, h_{zzz^*}, h_{zz^*z^*}) } \nl
  & \qquad =\:	& \bigg[	& \mc{2}{l}{ 1 - 3 \sigma^2 \la f''(H) \ra \frac1{\tilde{\sigma} \tilde{\tau}^2} \big( 2 \Re(h_{z^*z^*}h_{zz^*z^*}h_{zzz}) \big) } \nl
  &		&		& -\: \sigma^3 \la f'''(H) \ra \bigg(	& \frac{4}{\tilde{\tau}^2} - \frac{8|h_{zzz^*}|^2}{\tilde{\tau}^3} \nl
  &		&		&					& +\: \frac{2|h_{zzz^*}|^4 + 2 \Re(h_{zzz}h_{zz^*z^*}^3)}{\tilde{\tau}^4} \bigg) \bigg] \nl
  &		& \mc{3}{l}{ \times \frac1{\pi^3 \tilde{\sigma} \tilde{\tau}^2} \exp\Big( \!-\! \frac{|h_{zz}|^2}{\tilde{\sigma}} - \frac{|h_{zzz}|^2 + |h_{zzz^*}|^2}{\tilde{\tau}} \Big). }
\end{IEEEeqnarray}

		\section{Monstar fraction}
	\label{sec_monstarfr}

Once the joint probability distribution of the relevant derivatives is obtained, we can set $h_{zz} = h_{z^*z^*} = 0$, which defines an umbilical point. The joint probability distribution states how likely it is that $h_{zz}$ and $h_{z^*z^*}$ are \emph{close} to zero for a \emph{certain} point $\vec{r}$. What we need however is for $h_{zz}$ and $h_{z^*z^*}$ to be \emph{exactly} zero for a point \emph{close} to $\vec{r}$, since we are looking for a density with respect to the $(x,y)$-plane. For this, we need to go from a probability density with respect to $h_{zz}$ and $h_{z^*z^*}$ to one with respect to $z$ and $z^*$. This is accomplished by multiplying $p$ with the Jacobian
\begin{equation}
  J  =  \bigg| \frac{\partial(h_{zz}, h_{z^*z^*})}{\partial(z, z^*)} \bigg|  =  \big| |h_{zzz}|^2 - |h_{zzz^*}|^2 \big|.
  \label{eq_Jacobian}
\end{equation}

The last step is to integrate this product over $h_{zzz}$ and $h_{zzz^*}$, either over all possible values, or just over those satisfying eq.~\eqref{eq_monstar}, to get the density of umbilical points and the density of monstars respectively:
\begin{equation}
  \begin{split}
    n  =  \int_R \! \rd^2 h_{zzz} \rd^2 h_{zzz^*} \,	& p(h_{zz} = 0, h_{zzz}, h_{zzz^*}) \\
    							& \times J(h_{zzz}, h_{zzz^*}),
  \end{split}
\end{equation}
where $R$ represents the range of integration: the entire space to get the density of all umbilical points, or eq.~\eqref{eq_monstar} for just the monstars.

First we simplify by introducing polar coordinates,
\begin{align}
  h_{zzz}	& = |h_{zzz}|e^{i\phi},	& h_{zzz^*}	& = |h_{zzz^*}|e^{i\theta}.
\end{align}
Next, we introduce
\begin{align}
  u	& \equiv \frac{|h_{zzz}|}{|h_{zzz^*}|},	& \delta	& \equiv \frac{3\theta-\phi}2.
\end{align}
We find that we can rewrite the two conditions for monstars, eq.~\eqref{eq_monstar}, in terms of $u$ and $\delta$ only: the first one is simply $u > 1$, while the other is



\begin{align}
  0	& < 27 - u^4 - 18u^2 - 8u^3 \cos 2\delta \nl
  	& \quad = (3-u)^3(1+u) - 16u^3 \cos^2 \delta \nl
  	& \qquad \Leftrightarrow \cos^2 \delta < \frac{(3-u)^3(1+u)}{16u^3}.
  \label{eq_monstar2}
\end{align}
Since the fraction on the right-hand side is negative for $u > 3$, we can extend the first condition to $1 < u < 3$.

The fact that the monstar conditions depend only on $u$ and $\delta$ can be understood as follows: the type of umbilic should not be affected by rescaling and/or rotating the plane. Rescaling would add the same (real) factor to $h_{zzz}$ and $h_{zzz^*}$, hence the type of umbilic should, as far as the moduli are concerned, depend only on the ratio $|h_{zzz}| / |h_{zzz^*}|$. A rotation introduces phase factors as given by eq.~\eqref{eq_rotation}. We see that a rotation over an angle $\alpha$ causes $h_{zzz}$ to pick up a factor $e^{-3i\alpha}$ while $h_{zzz^*}$ picks up $e^{-i\alpha}$. Therefore, the only combination of $\phi$ and $\theta$ that is invariant under rotations is $3\theta-\phi$.

Now we return our attention to the probability distribution. First we rescale $h_{zzz}$ and $h_{zzz^*}$,
\begin{align}
  v	& \equiv \frac{h_{zzz}}{\sqrt{\tilde{\tau}}},	& w \equiv \frac{h_{zzz^*}}{\sqrt{\tilde{\tau}}}.
\end{align}
This leads to
\begin{IEEEeqnarray}{rll}
  & \mc{2}{l}{ p(h_{zz} = 0, v, w)J } \nl
  & \propto	& \Big( 1 - \frac{\sigma^3}{\tilde{\tau}^2} \la f'''(H) \ra (4 - 8|w|^2 + 2|w|^4 + vw^{*3} + v^*w^3) \Big)\nl
  &		& \times e^{-|v|^2-|w|^2} \big| |v|^2-|w|^2 \big|. \nonumber
\end{IEEEeqnarray}
Here we dropped an overall coefficient, which is of no importance since we are only interested in the ratio of the densities of monstars and all umbilical points. Note that $\tilde{\tau}$ now only appears in the term proportional to $f$. Since we are not interested in higher orders of $f$, we need only consider the leading order of $\tilde{\tau}$, which is $\tau$. For convenience, let us define
\begin{equation}
  \tilde{\eps}  \equiv  \frac{\sigma^3}{\tau^2} \la f'''(H) \ra.
\end{equation}
Furthermore, note that multiplying $p$ with the constant $1 + 4\tilde{\eps}$ -- which we may do since we are only interested in the density of the ratios -- causes the $4$ inside the parentheses to be canceled out (up to first order).

Next, we move to polar coordinates, as we did before: \footnote{Apart from the sign of $\phi$ and a numerical factor, the variables $r$, $\rho$, $\phi$ and $\theta$ match their respective counterparts in \cite{cite_Berry}.}
\begin{align}
  v	& \equiv \rho e^{i\phi},	& w	& \equiv  r e^{i\theta},
\end{align}
and then substitute $r=u\rho$ and $\phi=3\theta-2\delta$.
With these transformations we have
%
\begin{align}
  n \propto \int_R	& \rho^3u \, \rd \rho \rd u \rd \theta \rd \delta \, e^{-\rho^2(u^2+1)} \rho^2|u^2-1| \nl
  			& \! \times \big( 1 - \tilde{\eps} (-8\rho^2u^2 + 2\rho^4(u^4 + u^3\cos 2\delta) \big).
\end{align}
Finally, we integrate over $\rho$ and $\theta$ to find the probability distribution $p(u,\delta)$. The integration over $\theta$ simply gives a factor of $2\pi$, while the integral over $\rho$ has the form of a polynomial times a Gaussian. For this we can use
\begin{equation}
  \int_0^{\infty} \! \rd \rho \, \rho^{2n+1} e^{-\rho^2(u^2+1)}  =  \frac{n!}{2(u^2+1)^{n+1}}.
\end{equation}
The result is
\begin{equation}
  p(u,\delta) \propto \, \frac{u|u^2-1|}{(u^2+1)^3} \bigg( 1 + 24\tilde{\eps} \frac{u^2(1-u \cos 2\delta)}{(u^2+1)^2} \bigg).
  \label{eq_udelta}
\end{equation}

\begin{figure}
  \centering
  \includegraphics{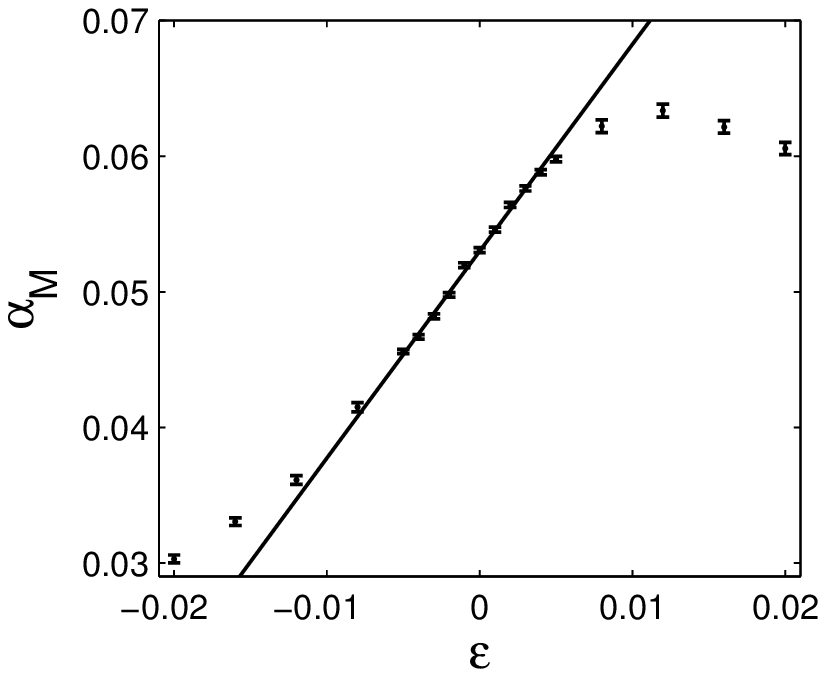}
  \caption{The monstar fraction $\alpha_M$ of $H + \eps H^3$ as a function of $\eps$, where $H$ has a disk spectrum ($\mu = \tfrac{16}{27}$). The data points stem from simulations, the solid line is eq.~\eqref{eq_mfr_cube}.}
  \label{fig_mfr_disk}
\end{figure}

\begin{figure}
  \centering
  \includegraphics{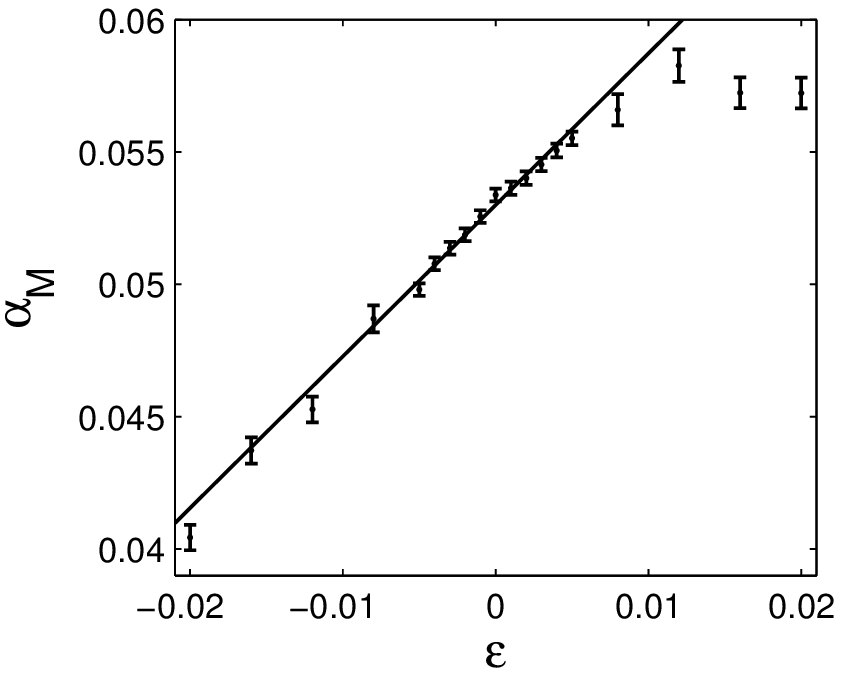}
  \caption{The monstar fraction $\alpha_M$ of $H + \eps H^3$ as a function of $\eps$, where $H$ has a Gaussian spectrum ($\mu = \tfrac29$). The data points stem from simulations, the solid line is eq.~\eqref{eq_mfr_cube}.}
  \label{fig_mfr_gauss}
\end{figure}

The monstar density is proportional to the integral of $p(u,\delta)$ over the range
\begin{equation}
  1 < u < 3,  \qquad  \cos^{-1} \Big( \sqrt{\frac{(3-u)^3(1+u)}{16u^3}} \Big) < \delta < \tfrac12\pi,
\end{equation}
while the total density of umbilical points is proportional (with the same prefactor) to the integral over the range $0 < u < \infty$, $0 < \delta < \tfrac12\pi$ (extending the integration range of $\delta$ from $\tfrac12\pi$ to $2\pi$ would just add a factor of $4$ to both integrals). The latter can be done analytically: the integration over $\delta$ is trivial, while the remaining integral over $\rho$ can be split into two parts which, apart from a factor $\sgn(u^2-1)$, are both of the form
\begin{equation}
  \int \! \rd u \, \frac{u^2-1}{(u^2+1)^2} \bigg( \frac{u}{u^2+1} \bigg)^n  =  -\frac1{n+1} \bigg( \frac{u}{u^2+1} \bigg)^{n+1}.
\end{equation}
With this we find
\begin{IEEEeqnarray}{rLl}
  n_{\mathrm{tot}}	& \propto	& \tfrac12\pi \bigg[ \tfrac12 \frac{u^2}{(u^2+1)^2} + 24\tilde{\eps} \tfrac14 \frac{u^4}{(u^2+1)^4} \bigg]_{u=0}^1 \nl
  			&		& +\: \tfrac12\pi \bigg[ \!-\! \tfrac12 \frac{u^2}{(u^2+1)^2} - 24\tilde{\eps} \tfrac14 \frac{u^4}{(u^2+1)^4} \bigg]_{u=1}^{\infty} \nl
  			& =		& \tfrac18\pi (1+3\tilde{\eps}).
\end{IEEEeqnarray}

For the monstar range, the integral over $\delta$ can be performed. The integral over the cosine gives
\begin{equation}
  \int_{\cos^{-1}(\ldots)}^{\tfrac12\pi} \! \rd \delta \, \cos 2\delta  =  -\frac1{16u^3} \sqrt{(u^2-1)(9-u^2)^3}.
\end{equation}
All together:
\begin{IEEEeqnarray}{rLlll}
  n_{\mathrm{M}}	& \propto	& \mc{3}{l}{ \int_1^3 \! \rd u \, \frac{u(u^2-1)}{(u^2+1)^3} \sin^{-1}\Bigg( \sqrt{\frac{(3-u)^3(1+u)}{16u^3}} \Bigg) } \nl
  			&		& +\: \tilde{\eps} \int_1^3 \!	& \mc{2}{l}{ \rd u \, \frac{u(u^2-1)}{(u^2+1)^3} } \nl
			&		&				& \times \Bigg[	& \frac{24u^2}{(u^2+1)^2} \sin^{-1}\Bigg( \sqrt{\frac{(3-u)^3(1+u)}{16u^3}} \Bigg) \nl
  			&		&				&		& +\: \frac{3\sqrt{(u^2-1)(9-u^2)^3}}{2(u^2+1)^2} \Bigg] \nl
  			& \mc{4}{L}{ \equiv I_1 + \tilde{\eps} I_2. }
\end{IEEEeqnarray}

The integrals can be done numerically. The monstar fraction is then
\begin{align}
  \alpha_M	& = \frac{n_{\mathrm{M}}}{n_{\mathrm{tot}}}  =  \frac{I_1 + \tilde{\eps} I_2}{\tfrac18\pi (1+3\tilde{\eps})}  =  \frac{8}{\pi}(I_1 + \tilde{\eps}(I_2 - 3I_1)) + O(\tilde{\eps}^2) \nl
  		& = 0.053 + 0.429 \mu \la f'''(H) \ra + O(f^2),
  \label{eq_mfr}
\end{align}
where
\begin{equation}
  \mu  \equiv  \frac{\sigma^3}{\tau^2}  =  \frac{K_4^3}{K_6^2}  \qquad  (0 \leq \mu \leq 1).
\end{equation}
Note that the zeroth order result matches the one in \cite{cite_Berry}.

Remember that we set $K_0 = \la H^2 \ra = 1$ for convenience; if we drop this condition, then $K_0$ enters the denominator of the expression above.

There is an alternative expression for the term $\la f'''(H) \ra$ in eq.~\eqref{eq_mfr}. Since $H$ is Gaussian, with mean $0$ and deviation $K_0 = 1$, we can write
\begin{equation}
  \la f'''(H) \ra  =  \int \! \rd z \, f'''(z) e^{-z^2/2}.
\end{equation}
Repeated partial integration yields
\begin{align}
  \la f'''(H) \ra	& = \int \! \rd z \, z f''(z) e^{-z^2/2} \nl
  			& = \int \! \rd z \, (z^2-1) f'(z) e^{-z^2/2} \nl
  			& = \int \! \rd z \, (z^3-3z) f(z) e^{-z^2/2} \nl
  			& = \la (H^3-3H) f(H) \ra.
\end{align}
The \emph{kurtosis} of a stochastic variable is defined as the fourth cumulant divided by the square of the second, which gives
\begin{equation}
  \kappa  \equiv  \frac{ \la h^4 \ra }{ \la h^2 \ra ^2 } - 3.
\end{equation}
If we enter $h = H + f(H)$, we find
\begin{align}
  \kappa	& = \frac{ \la H^4 \ra + 4 \la H^3 f(H) \ra }{ \la H^2 \ra + 4 \la H f(H) \ra } - 3 + O(f^2) \nl
  		& = \frac{ 4 \la H^3 f(H) \ra - 12 \la H f(H) \ra }{ 1 + 4 \la H f(H) \ra } + O(f^2) \nl
  		& = 4 \la H^3 f(H) \ra - 12 \la H f(H) \ra.
\end{align}
We see that $\la f'''(H) \ra = \kappa / 4$ up to first order, which remains true if $K_0 \neq 1$. Hence an alternative form of eq.~\eqref{eq_mfr} is
\begin{equation}
  \alpha_M  =  0.053 + 0.107 \mu \kappa + O(f^2),
\end{equation}
where $\kappa$ is the kurtosis of $h$.

By comparing the fraction of monstars in a given field to the formula just found, we can determine one parameter of the deviation from a Gaussian distribution, $\langle f'''(H)\rangle$. This assumes that the field $h$ is given by $h = H + f(H)$. To test this, one could, if possible, also measure the distribution $p(u,\delta)$ to test that it has the right form, eq.~\eqref{eq_udelta}. Measuring $p(\delta)$ only could also suffice. For a Gaussian field $H$, all values of $\delta$ should be equally likely, whereas integrating eq.~\eqref{eq_udelta} shows that the distribution we expect for $h$ is
\begin{equation}
  p(\delta)  =  \frac1{\pi} (1-4\tilde{\eps}\cos 2\delta),
  \label{eq_delta}
\end{equation}
where we define $\delta$ to lie between $-\pi/2$ and $\pi/2$.

		\section{Comparison with simulations}
	\label{sec_sims}

The most basic example of a non-Gaussian variable for which eq.~\eqref{eq_mfr} can be tested is $h = H + \eps H^3$, for which $\la f'''(H) \ra = 6\eps$. Then we have
\begin{equation}
  \alpha_M  =  0.053 + 2.576\mu\eps + O(\eps^2).
  \label{eq_mfr_cube}
\end{equation}
Eq.~\eqref{eq_mfr} was compared to results from simulations. Gaussian fields $H$ of square size were generated by adding together a large number (a few hundred) of waves with random phases, in the spirit of eq.~\eqref{eq_gsn_field}. We then counted the number of monstars and umbilics in $h = H + \eps H^3$ for various values of $\eps$. Periodic boundary conditions were applied to reduce finite-size effects. Furthermore, we chose spectra for which the spectrum decays very quickly (or is zero) for large $k$: a disk spectrum,
\begin{equation}
  A(k)^2 \sim \theta(k - k_0)  \qquad  K_{2n} = \frac{k_0^{2n}}{n+1}  \qquad  \mu = \frac{16}{27},
\end{equation}
and a Gaussian spectrum,
\begin{equation}
  A(k)^2 \sim \exp(-k^2 / 2k_0^2)  \qquad  K_{2n} = 2^n n! k_0^{2n}  \qquad  \mu = \frac29.
\end{equation}
The simulations were done in the same way as described in \cite{cite_paper1}.

A very good agreement between theory and simulation was found for both spectra (see figs.~\ref{fig_mfr_disk} and \ref{fig_mfr_gauss}), for $\eps$ up to about $0.01$. For larger values of $\eps$, nonlinear terms start to dominate.

Another thing to note is the sensitivity: in eq.~\eqref{eq_mfr_cube} we see that the prefactor of the perturbation term is very large compared to the leading order. As a result, even for small $\eps$, the relative deviation from the universal $0.053$ is quite large, as can be seen in the graphs. Therefore, measuring the monstar fraction of a given field proves to be a good method for detecting and quantifying small deviations from Gaussianity.

		\section{Conclusion}
	\label{sec_concl}

We have calculated how the density of monstars changes for a non-Gaussian field given by $h = H + f(H)$, where $H$ is a Gaussian field. Comparing our formula to data allows to measure the parameter $\langle f'''(H)\rangle = \kappa/4$ of the non-Gaussian contribution. Furthermore, one can also measure the distribution of the parameters $u$ and $\delta$ that define the type of umbilical point. The expected distribution is eq.~\eqref{eq_udelta}, or eq.~\eqref{eq_delta} for just the variable $\delta$. Measuring these distributions further constrains the value of $\la f'''(H)\ra$, and more importantly, it gives a test that the non-Gaussianity really arises from a local nonlinear transformation of a Gaussian field.


Even though in general the derivatives of $h$ have an infinite number of nonzero cumulants up to first order in $f$, it turns out that the cumulants of the variables which are relevant for the umbilics ($h_{zz}$, $h_{zzz}$ and $h_{zzz^*}$ at a single point) vanish beyond the fourth order due to symmetry. As a result, we found the interesting result that (up to first order) $\alpha_M$ depends only on $\la f'''(H) \ra$.

For a more general type of non-Gaussian field there would be more independent variables and hence more nonzero cumulants. However, it often still holds that the higher order cumulants are of less importance. In this case, one can consider only the cumulants up to a specific order (e.g.\ fourth order), of which still many would be zero due to symmetry. Applying the same procedure as outlined here could then reveal the monstar fraction up to first order.

\begin{acknowledgments}
This work was supported by the Dutch Foundation for Fundamental Research on Matter (FOM), the Dutch Foundation for Scientific Research (NWO) and the European Research Council (ERC).
We thank T. Lubensky, R.D. Kamien, B. Jain, Alexey Boyarsky, L. Mahadevan, B. Chen and W. van Saarloos for stimulating discussions.
\end{acknowledgments}

\appendix

\section{Proof of eq.~\eqref{eq_cum_with_f}}
\label{app_cum_with_f}

In this section we prove the identity
\begin{equation}
  \begin{split}
    & C_n(f(H_1), H_2, \ldots, H_n) \\
    & \qquad = \la f^{(n-1)}(H_1) \ra \la H_1H_2 \ra \la H_1H_3 \ra \ldots \la H_1H_n \ra,
  \end{split}
\end{equation}
for Gaussian variables $H_i$.
  
Recall the definition of a cumulant, eq.~\eqref{eq_log_chi}. The generating function $\chi$ is the Fourier transform of the probability distribution (see eq.~\eqref{eq_gen_func}), which for Gaussian variables is eq.~\eqref{eq_gauss_distr}. This leads to the identity

\begin{widetext}
  \begin{equation}
    \begin{split}
      & C_n(f(H_1), H_2, \ldots, H_n) \\
      & \qquad = (-i)^n \frac{\partial}{\partial \lambda_1} \ldots \frac{\partial}{\partial \lambda_n} \log \int \rd h_1 \ldots \rd h_n \,
        e^{i(\lambda_1 f(h_1) + \lambda_2 h_2 + \ldots + \lambda_n h_n)} \frac{ \exp \big( \!-\! \tfrac12 \sum_{ij} \sigma^{-1}_{ij} h_i h_j \big) }{ (2\pi)^{n/2} \sqrt{\det \sigma} } \, \Bigg|_{\lambda_1 = \ldots = \lambda_n = 0}.
    \end{split}
    \label{eq_tocalculate}
  \end{equation}
\end{widetext}

First, the integration over $h_2$ through $h_n$ is performed. This partial Fourier transform is not trivial. If $h_1$ were included, the answer would be simply eq.~\eqref{eq_chi_gauss}. We can however use this result and take the inverse Fourier transform of it with respect to $\lambda_1$ to get the desired result:
\begin{IEEEeqnarray}{llll}
  & \mc{3}{l}{ \int \rd h_2 \ldots \rd h_n \, e^{i(\lambda_2 h_2 + \ldots + \lambda_n h_n)} \frac{ \exp \big( \!-\! \frac12 \sum_{ij} \sigma^{-1}_{ij} h_i h_j \big) }{ (2\pi)^{n/2} \sqrt{\det \sigma} } } \nl
  & \mc{3}{l}{ \quad = \int \frac{\rd \lambda_1}{2\pi} e^{-i \lambda_1 h_1} \exp \Big( \!-\! \tfrac12 \sum_{ij} \sigma_{ij} \lambda_i \lambda_j \Big) } \nl
  & \quad = \int \frac{\rd \lambda_1}{2\pi} \exp \Big[	& \mc{2}{l}{ - \tfrac12 \sigma_{11} \lambda_1^2 - \big( i h_1 + \sum_{j\geq2} \sigma_{1j} \lambda_j \big) \, \lambda_1 } \nl
  &							& \mc{2}{l}{ \:-\: \tfrac12 \sum_{i,j\geq2} \sigma_{ij} \lambda_i \lambda_j \Big] } \nl
  & \mc{2}{l}{ \quad = \frac1{\sqrt{2\pi \sigma_{11}}} \exp \Big[\! }	& \frac1{2\sigma_{11}} \big( i h_1 + \sum_{j\geq2} \sigma_{1j} \lambda_j \big)^2 \nl
  &							&		& \:-\: \tfrac12 \sum_{i,j\geq2} \sigma_{ij} \lambda_i \lambda_j \Big].
\end{IEEEeqnarray}
The integration was performed by completing the square.
Now we do the Fourier transform with respect to $h_1$ and include the logarithm present in eq.~\eqref{eq_tocalculate}, which leads to
\begin{IEEEeqnarray}{lllll}
  & \mc{2}{l}{ \log \int \rd h_1 \ldots \rd h_n }	& \mc{2}{l}{ e^{i(\lambda_1 f(h_1) + \lambda_2 h_2 + \ldots + \lambda_n h_n)} } \nl
  &			&				& \mc{2}{l}{ \frac{ \exp \big( \!-\! \tfrac12 \sum_{ij} \sigma^{-1}_{ij} h_i h_j \big) }{ (2\pi)^{n/2} \sqrt{\det \sigma} } } \nl
  & \: = \log \int	& \mc{3}{l}{ \rd h_1 e^{i \lambda_1 f(h_1)} \frac1{\sqrt{2\pi \sigma_{11}}} } \nl
  & 			& \mc{3}{l}{ e^{ \frac1{2\sigma_{11}} [ i h_1 + \sum_{j\geq2} \sigma_{1j} \lambda_j ]^2 - \frac12 \sum_{i,j\geq2} \sigma_{ij} \lambda_i \lambda_j } } \nl
  & \mc{3}{l}{ \: = -\tfrac12 \sum_{i,j\geq2} \sigma_{ij} \lambda_i \lambda_j + \log \int \! }	& \rd h_1 e^{i \lambda_1 f(h_1)} \frac1{\sqrt{2\pi \sigma_{11}}} \nl
  &			&				&					& e^{ \frac1{2\sigma_{11}} [ i h_1 + \sum_{j\geq2} \sigma_{1j} \lambda_j ]^2 }
\end{IEEEeqnarray}
In accordance with eq.~\eqref{eq_tocalculate}, we must take the derivative of this equation with respect to the $\lambda$'s and set them to zero. First the derivative with respect to $\lambda_1$ is taken. This causes the first term to vanish, since it does not depend on $\lambda_1$. This simplification can be regarded as the main reason why the final result depends on $f$ in a rather simple way. What remains is
\begin{IEEEeqnarray}{lll}
  & -i \frac{\partial}{\partial \lambda_1} \log \int	& \rd h_1 e^{i \lambda_1 f(h_1)} \frac1{\sqrt{2\pi \sigma_{11}}} \nl
  &							& e^{ \frac1{2\sigma_{11}} [ i h_1 + \sum_{j\geq2} \sigma_{1j} \lambda_j ]^2 } \bigg|_{\lambda_1=0} \nl
  & \mc{2}{l}{ \quad = \frac	{ \int \rd h_1 \, f(h_1) \exp \big( \frac1{2\sigma_{11}} \big( i h_1 + \sum_{j\geq2} \sigma_{1j} \lambda_j \big)^2 \big) }
				{ \int \rd h_1 \, \exp \big( \frac1{2\sigma_{11}} \big( i h_1 + \sum_{j\geq2} \sigma_{1j} \lambda_j \big)^2 \big) } } \nl
  & \mc{2}{l}{ \quad = \frac1{\sqrt{2\pi \sigma_{11}}} \int \rd h_1 \, f(h_1) e^{ \frac1{2\sigma_{11}} [ i h_1 + \sum_{j\geq2} \sigma_{1j} \lambda_j ]^2 }. }
\end{IEEEeqnarray}
For each $\lambda_k$ with $k \geq 2$ the derivative yields
\begin{equation}
  \begin{split}
    & -i \frac{\partial}{\partial \lambda_k} \, e^{ \frac1{2\sigma_{11}} [ i h_1 + \sum_{j\geq2} \sigma_{1j} \lambda_j ]^2 } \\
    & \quad = -i \frac{\sigma_{1k}}{\sigma_{11}} \big(i h_1 + \sum_{j\geq2} \sigma_{1j} \lambda_j \big) e^{ \frac1{2\sigma_{11}} [ i h_1 + \sum_{j\geq2} \sigma_{1j} \lambda_j ]^2 }.
  \end{split}
\end{equation}
Applying this for all $k$ and subsequently setting all $\lambda_k$ to zero then gives
\begin{equation}
  \begin{split}
    & (-i)^{n-1} \frac{\partial}{\partial \lambda_2} \ldots \frac{\partial}{\partial \lambda_n} \, e^{ \frac1{2\sigma_{11}} [ i h_1 + \sum_{j\geq2} \sigma_{1j} \lambda_j ]^2 } \\
    & \qquad = \bigg( \frac{h_1}{\sigma_{11}} \bigg)^{n-1} \bigg( \prod_{j\geq2} \sigma_{1j} \bigg) e^{-h_1^2/(2\sigma_{11})}.
  \end{split}
\end{equation}
This results in
\begin{equation}
  \begin{split}
    & C_n(f(H_1), H_2, \ldots, H_n) = \bigg( \prod_{j\geq2} \sigma_{1j} \bigg) \\
    & \qquad \times \frac1{\sqrt{2\pi \sigma_{11}}} \int \rd h_1 \, f(h_1) \bigg( \frac{h_1}{\sigma_{11}} \bigg)^{n-1} e^{-h_1^2/(2\sigma_{11})}.
  \end{split}
\end{equation}
Integrating by parts $n-1$ times leads to
\begin{equation}
  \begin{split}
    & C_n(f(H_1), H_2, \ldots, H_n) = \bigg( \prod_{j\geq2} \sigma_{1j} \bigg) \\
    & \qquad \times \frac1{\sqrt{2\pi \sigma_{11}}} \int \rd h_1 \, f^{(n-1)}(h_1) e^{-h_1^2/(2\sigma_{11})}.
  \end{split}
\end{equation}
Finally, we identify the integral (along with the prefactor) as the expectation value of $f^{(n-1)}$ and $\sigma_{1j} = \la H_1 H_j \ra$, which gives us
\begin{equation}
  \begin{split}
    & C_n(f(H_1), H_2, \ldots, H_n) \\
    & \qquad = \la f^{(n-1)}(H_1) \ra \la H_1H_2 \ra \la H_1H_3 \ra \ldots \la H_1H_n \ra,
  \end{split}
\end{equation}
the equation we set out to prove.

\bibliography{nongauss}

\end{document}